\documentclass{research4cacm}
\usepackage{textcomp}
\usepackage{url}
\usepackage{listings}
\usepackage{pgfplots}
\usepackage{subfig}
	\usepackage{fixltx2e} 
	\usepackage{xcolor}
	\definecolor{ForestGreen}{rgb}{0.0, 0.4, 0.0}
	\colorlet{HeadingColor}{ForestGreen}
	\definecolor{NavyBlue}{rgb}{0.0, 0.0, 0.5}
	
	\definecolor{darkmagenta}{rgb}{0.55, 0.0, 0.55}
	\definecolor{burntorange}{rgb}{0.8, 0.33, 0.0}
	\definecolor{ivory}{rgb}{1.0, 1.0, 0.94}
   \definecolor{webyellow}{rgb}{0.98,0.92,0.73}
 \colorlet{SeparatorColor}{HeadingColor}
\usepackage{adjustbox}
\usepackage{tikz}
\usetikzlibrary{arrows,shadows,backgrounds,calc,automata,
decorations.pathreplacing,decorations.pathmorphing,shapes.arrows}
\usetikzlibrary{decorations.markings}
\definecolor{burntorange}{rgb}{0.8, 0.33, 0.0}
\definecolor{ForestGreen}{rgb}{0.0, 0.4, 0.0}
\definecolor{NavyBlue}{rgb}{0.0, 0.0, 0.5}
\colorlet{Color1}{ForestGreen}
\colorlet{Color2}{burntorange}
\colorlet{Color3}{NavyBlue}
	\definecolor{webgreen}{rgb}{0,.5,0}
	\definecolor{webbrown}{rgb}{.6,0,0}
	\definecolor{webyellow}{rgb}{0.98,0.92,0.73}
	\definecolor{webgray}{rgb}{.753,.753,.753}
	\definecolor{webblue}{rgb}{0,0,.8}
    \definecolor{webgreen}{rgb}{0, 0.7, 0} 
    \definecolor{webred}{rgb}{0.8, 0, 0}   

\pgfplotscreateplotcyclelist{my color list}{%
solid, color=webblue, every mark/.append style={solid, fill=webblue}, mark=*\\%
densely dashdotted, color=webgreen, every mark/.append style={solid, fill=webred},mark=diamond*\\%
densely dotted, color=webbrown, every mark/.append style={solid, fill=webgreen}, mark=triangle*\\%
loosely dashed, color=webred, every mark/.append style={solid, fill=webbrown},mark=*\\%
dotted, color=webblue, every mark/.append style={solid, fill=webyellow}, mark=square*\\%
densely dotted, color=webgreen, every mark/.append style={solid, fill=gray}, mark=otimes*\\%
dashed, color=webbrown, every mark/.append style={solid, fill=gray},mark=diamond*\\%
densely dashed, every mark/.append style={solid, fill=gray},mark=square*\\%
dashdotted, every mark/.append style={solid, fill=gray},mark=otimes*\\%
dasdotdotted, every mark/.append style={solid},mark=star\\%
}

\newlength{\nd}	
\makeatletter
\newcommand{\gettikzxy}[3]{%
  \tikz@scan@one@point\pgfutil@firstofone#1\relax
  \edef#2{\the\pgf@x}%
  \edef#3{\the\pgf@y}%
}
\makeatother
\begin{document}
%
\CopyrightYear{2015} 

\title{
Can Broken Multicore Hardware be Mended?}
\numberofauthors{1} 
\author{
\alignauthor
J\'anos V\'egh\\
       \affaddr{University of Miskolc, Hungary}\\
       \affaddr{Department of Mechanical Engineering and Informatics}\\
       \affaddr{3515 Miskolc-University Town, Hungary}\\
       \email{J.Vegh@uni-miskolc.hu}
}

\maketitle
\begin{abstract}
A suggestion is made for mending multicore hardware, which has been diagnosed as broken.

\end{abstract}

	\lstset{
		 numbers=left,               
		numberstyle=\tiny,          
		numbersep=5pt,              
		tabsize=2,                 	 
		inputencoding=utf8/latin2,	
		extendedchars=true,         %
		escapechar=\@,
		breaklines=true,        
		columns=fullflexible,  
		breakatwhitespace=true,    
		escapeinside={\%*}{*)},          
		frame=tb, 
		framerule=.5pt, 
		rulecolor= \color{SeparatorColor},
		backgroundcolor=\color{ivory},
		basicstyle=\ttfamily\color{black}\lstsize\bfseries, 
		keywordstyle=\bfseries\color{darkmagenta},
		identifierstyle=\bfseries\color{NavyBlue},
		commentstyle=\itshape\bfseries\color{ForestGreen},
		stringstyle=\itshape\bfseries\color{burntorange}, 
		lineskip=0pt,aboveskip=4pt,belowskip=2pt,
		framesep=4pt,rulesep=2pt, 
		showspaces=false,           %
		showtabs=false,             %
		framexleftmargin=0pt,
		framexrightmargin=0pt,
		showstringspaces=false      
	}	
	\lstset{language={[ANSI]C}, basicstyle=\ttfamily\color{black}\normalsize}

\section{The multicore era is a consequence of the stalling of the
single-thread performance}\label{sec:introduction}

The multi- and many-core  (MC) era we have reached was triggered after the beginning of the century by the stalling of 
single-processor performance.  
Technology allowed more transistors to be placed on a die, but 
they could not reasonably be utilized to increase single-processor performance.
Predictions 
about the number of cores
has only partly been fulfilled: today's processors have dozens rather than the predicted hundreds of cores
(although the Chinese supercomputer~\cite{FuSunwaySystem2016} announced in the middle of 2016 comprises 260 cores on a die).
Despite this, the big players are optimistic.
They expect that Moore-law persists, 
though based on presently unknown technologies.
The effect of the stalled clock frequency is mitigated,
and it is even predicted~\cite{Intel10GHz:2014} that "\textit{Now that there are multicore processors, there is no reason why computers shouldn't begin to work faster, whether due to higher frequency or because of parallel task execution. And with parallel task execution it provides even greater functionality and flexibility!.}"

Parallelism is usually considered in many forums~\cite{ComputingPerformance:2011} to be the future, usually as the 
only hope, rather than as a panacea. 
People dealing with parallelism are less optimistic.
In general, the technical development tends to reduce the human effort,
but  "\emph{parallel programs ... are notoriously difficult to write, test, analyze, debug, and verify, much more so than the sequential versions}"~\cite{ReliableParallel2014}.
The problems have led researchers to the \textit{ViewPoint}~\cite{Vishkin:BrokenManycoreCACM}, that 
\emph{multicore hardware for general-purpose parallel processing is broken}.

\section{Manycore architectures could be fresh meat on the market of processors, but they are not}\label{sec:freshmeat}

The essence of the present Viewpoint is that multicore hardware can perhaps be mended. Although one can
profoundly agree with the arguments~\cite{Vishkin:BrokenManycoreCACM}
that using manycore chips cannot contribute much to using parallelism in general,
and especially not in executing irregular programs, one has to realize 
also that this is not the optimal battlefield for the manycore chips,
at least not in their present architecture.
Present manycore systems comprise many segregated processors,
which make no distinction between two processing units that are neighbours within
the same chip or  are located in the next rack. The close physical proximity
of the processing units offers additional possibilities, and provides a chance
to implement Amdahl's dream~\cite{AmdahlSingleProcessor67} of cooperating processors.

Paradigms  used presently, however, assume a private processor
and a private address space for a running process, and no external world.
In many-core systems, it is relatively simple to introduce signals,
storages, communication, etc., and deploy them in reasonable times.
They cannot, however, be utilized in a reasonable way, if one cannot provide
compatibiliy facades providing the illusion of the private world.
Cooperation must be implemented in a way which provides
complete (upward) compatibility with the presently exclusively used Single-Processor Approach~(SPA)~\cite{AmdahlSingleProcessor67}.
It means that on the one hand that new functionality must be formulated using the
terms of conventional computing, while on the other, it
provides considerably enhanced computing throughput and other advantages.

It is well known, that general purpose processors 
have a huge handicap in performance when compared to  
special purpose chips, and that the presently used computing stack
is the source of further serious inefficiencies. 
Proper utilization of available manycore processors
can eliminate a lot of these performance losses,
and in this way (keeping the same electronic and programming technology) 
can considerably enhance (apparently) the performance of 
the processor. Of course, there is no free lunch. Making these changes 
requires a \textit{simultanous} change in nearly all elements
of the present computing stack.
Before making these changes, one should scrutinize the promised gain,
and whether the required efforts will pay off.

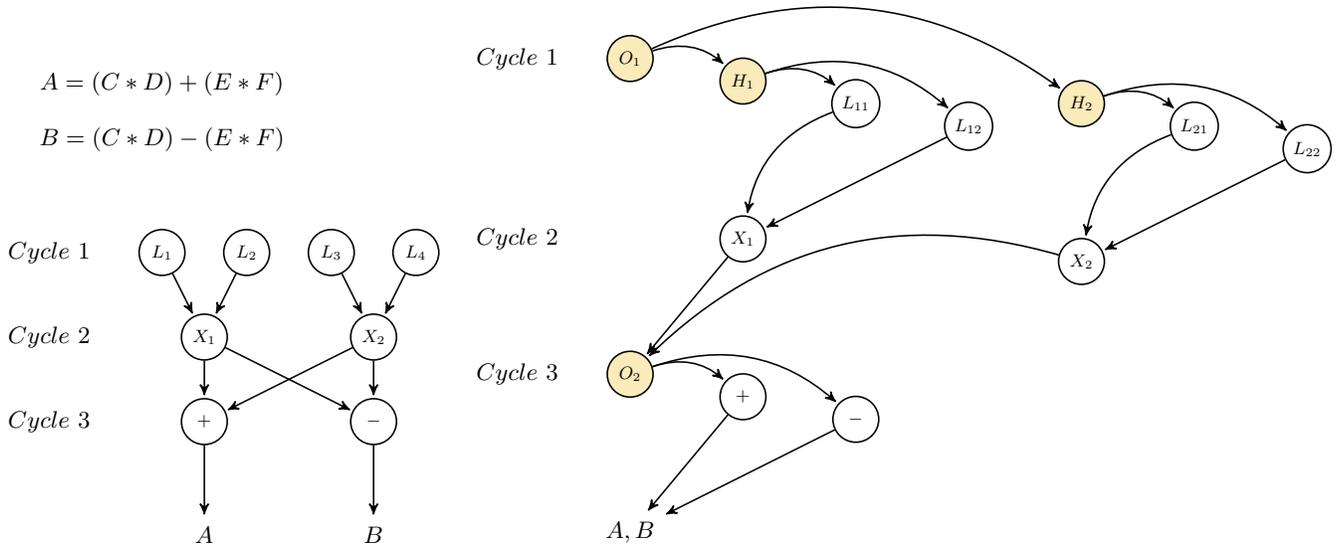
\begin{figure*}
\begin{tabular}{cc}
\begin{tikzpicture}[->,>=stealth',shorten >=1pt,auto,node distance=\nd and \nd,
                    semithick]
  \tikzstyle{every state}=[scale=0.75]
  \node[state]  (L1)       {$L_1$};
  \node         (E1) [ above of=L1]      {$B = (C*D)-(E*F)$};
  \node         (E2m) [ above of=E1]      {};
  \node         (E2) at ($(E1)!0.5!(E2m)$)     {$A = (C*D)+(E*F)$};

  \node[state]  (L2) [ right of=L1]     {$L_2$};
  \node[state]  (L3) [ right of=L2]     {$L_3$};
  \node[state]  (L4) [ right of=L3]     {$L_4$};

  \node  (X1F)  at ($(L1)!0.5!(L2)$)     {};
  \node[state]  (X1)   [ below of=X1F]     {$X_1$};
  \node  (X2F)  at ($(L3)!0.5!(L4)$)     {};
  \node[state]  (X2)   [ below of=X2F]     {$X_2$};
 
  \node[state]  (plus) [ below of=X1]     {$+$};
  \node[state]  (minus) [ below of=X2]     {$-$};
  
  \node (A) [ below of=plus]     {$A$};
  \node (B) [ below of=minus]     {$B$};

 \path 
   (L1) edge  (X1)
   (L2) edge  (X1)
   (L3) edge  (X2)
   (L4) edge  (X2);
 \path 
   (X1) edge  (plus)
   (X1) edge  (minus)
   (X2) edge  (plus)
   (X2) edge  (minus);
 \path 
   (plus) edge  (A)
   (minus) edge  (B);

  \node         (C1) [ left of=L1] {$Cycle\ 1$};
\gettikzxy{(C1)}{\cx}{\cy}
\gettikzxy{(X1)}{\xx}{\xy}
\gettikzxy{(plus)}{\px}{\py}

  \node         (C2) at ($(\cx,\xy)$) {$Cycle\ 2$};
  \node         (C3) at ($(\cx,\py)$) {$Cycle\ 3$};

\end{tikzpicture}
&
\begin{tikzpicture}[->,>=stealth',shorten >=1pt,auto,
node  distance=\nd and \nd,
                    semithick]
  \tikzstyle{every state}=[scale=0.75]
  
  \node[state,fill=webyellow]  (O1)                    {$O_1$};
  \node[state,fill=webyellow]  (H1) at ($(O1)+(\nd,-.2\nd)$)       {$H_1$};
  \node[state,fill=webyellow]  (H2) at ($(O1)+(4\nd,-.4\nd)$)       {$H_2$};
   \path 
     (O1) edge [bend left] node {} (H1)
     (O1) edge [bend left] node {} (H2);
  
  \node[state]  (L11) at ($(H1)+(\nd,-.2\nd)$)       {$L_{11}$};
  \node[state]  (L12) at ($(H1)+(2\nd,-.4\nd)$)       {$L_{12}$};
   \path 
     (H1) edge [bend left] node {} (L11)
     (H1) edge [bend left] node {} (L12);

  \node[state]  (L21) at ($(H2)+(\nd,-.2\nd)$)       {$L_{21}$};
  \node[state]  (L22) at ($(H2)+(2\nd,-.4\nd)$)       {$L_{22}$};
   \path 
     (H2) edge [bend left] node {} (L21)
     (H2) edge [bend left] node {} (L22);
  
  \node[state]  (X1) at ($(H1)+(0,-1.4\nd)$)       {$X_1$};
   \path 
     (L11) edge [bend right] node {} (X1)
     (L12) edge   node {} (X1);
  \node[state]  (X2) at ($(H2)+(0,-1.4\nd)$)       {$X_2$};
   \path 
     (L21) edge [bend right] node {} (X2)
     (L22) edge   node {} (X2);

\gettikzxy{(O1)}{\ox}{\oy}
\gettikzxy{(X2)}{\xx}{\xy}
  \node[state,fill=webyellow]  (O2)   at ($(\ox,\xy)-(0,\nd)$)                 {$O_2$};
   \path 
     (X1) edge  node {} (O2)
     (X2) edge  [bend right]   node {} (O2);

  \node[state]  (plus) at ($(O2)+(\nd,-.2\nd)$)       {$+$};
  \node[state]  (minus) at ($(O2)+(2\nd,-.4\nd)$)       {$-$};
   \path 
     (O2) edge [bend left] node {} (plus)
     (O2) edge [bend left] node {} (minus);

\gettikzxy{(minus)}{\minx}{\miny}
  \node  (O3)   at ($(\ox,\miny)-(0,\nd)$)                 {$A,B$};
   \path 
     (plus) edge  node {} (O3)
     (minus) edge   node {} (O3);
     
  \node         (C1) [ left of=O1] {$Cycle\ 1$};
\gettikzxy{(C1)}{\cx}{\cy}
\gettikzxy{(X1)}{\xx}{\xy}
  \node         (C2) at ($(\cx,\xy)$) {$Cycle\ 2$};
\gettikzxy{(O2)}{\px}{\py}
  \node         (C3) at ($(\cx,\py)$) {$Cycle\ 3$};
\end{tikzpicture}
\\
\end{tabular}
\caption{Theoretical parallelism (left) vs dynamic parallelism implemented on a processor system with runtime configurable  architecture (right).}
\label{fig:flexibleproc}
\end{figure*}

Below, some easy-to follow case studies are presented, all of which
 lead to the same conclusion:
 we need a cooperative and flexible rather than rigid architecture comprising segregated  MCs, and the 70-years-old von Neumann computing paradigms should be extended. At the end, the feasibility of implementing such an
 architecture is discussed. 
 The recently introduced Explicitly Many-Processor Approach~\cite{VeghDynamicParallelism:2016} seems to be quite promising:
 it not only provides higher computing throughput, but also offers
 advantageous changes in the behavior of computing systems.

\section{Is implementing mathematical parallelism just a dream?}\label{sec:manymany}

Todays computing utilizes many forms of parallelism~\cite{HwangParallelism:2016}, both hardware (HW) and software (SW) facilities.
The software is systematically discussed in~\cite{Vishkin:BrokenManycoreCACM}
and hardware methods are scrutinized in~~\cite{HwangParallelism:2016}.
A remarkable difference between the two approaches is, that while 
the SW methods tend to handle the parallel execution explicitly,
the HW methods tend to create the illusion that only one processing unit can
cope with the task, although some (from outside invisible) helper units
are utilized in addition to the visible processing unit. 
Interestingly enough, both approaches arise from the von Neumann paradigms:
the abstractions \textit{process} and the \textit{processor} require so.

The inefficiency of using several processing units is nicely illustrated
with a simple example in~\cite{HwangParallelism:2016} (see also Fig~\ref{fig:flexibleproc}, left side).
A simple calculation comprising 4 operand loadings and 4 aritmetic operations,
i.e. altogether  8 machine instructions, could be theoretically carried out in
3 clock cycles, provided that only dependencies restrict the execution of 
the instructions and an unlimited number of processing units (or at least 4 such units in the example) are available.
It is shown that a single-issue processor needs 8 clock cycles to carry out
the calculation example.

Provided that memory access and instruction latency time cannot be
further reduced, the only possibility to shorten execution time is to use more than one processing unit
during the calculation. Obviously, a fixed architecture can only provide
a fixed number of processing units. In the example~\cite{HwangParallelism:2016}
two such ideas are scrutinized: a dual-issue single processor, and a two-core single issue
 processor. The HW investment in both cases increases by a factor of two (not considering the shared memory here), while the performance increases
only moderately: 7 clock cycles for the dual-issue processor and 6 clock cycles
for the dual-core processor, versus the 8 clock cycles of the single-issue
single core processor. The \textit{obvious reasons here are the rigid architecture and the lack of communication possibilities}, respectively.

Consider now a processor with flexible architecture, where the processor
can outsource part of its job: it can rent processing units from a chip-level pool just in the time it takes to execute a few instructions.
The cores are smart: they can communicate with each other, and even they know the
task to be solved and are able to organize their own work while outsourcing part of the work to the rented cores.
The sample calculation, borrowed from~\cite{HwangParallelism:2016}
as shown in Fig.~\ref{fig:flexibleproc}, left side,
can then be solved as shown on the right side of the figure.

\def\corescale{0.5}
The core
\tikz \node[state,scale=\corescale] {$O_1$};  
originally receives the complete task to make the calculation,
as it would be calculated by a conventional single-issue, single core system, in 8 clock cycles.
However,
\tikz \node[state,scale=\corescale]       {$O_1$};
is more intelligent. Using the hints hidden in the object code, it notices that the task can be outsourced
to another cores. For this purpose it rents, one by one, cores
\tikz \node[state,scale=\corescale]  {$H_1$};  and \tikz \node[state,scale=\corescale] {$H_2$};
to execute two multiplications. The rented
\tikz \node[state,scale=\corescale] {$H_x$};
cores are also intelligent, so they also outsource loading the operands to cores
\tikz \node[state,scale=\corescale] {$L_{x1}$}; 
and 
\tikz \node[state,scale=\corescale] {$L_{x2}$};. 
They execute the outsourced job: load the operands and return them to
the requesting cores 
\tikz \node[state,scale=\corescale] {$H_x$};,
which then can execute the multiplications
(denoted by
\tikz \node[state,scale=\corescale] {$X_x$};) and return the result to the requesting core, which can 
then rent another two cores
\tikz \node[state,scale=\corescale] {$+$};  and 
\tikz \node[state,scale=\corescale] {$-$}; for the final operations.
Two results are thus produced.

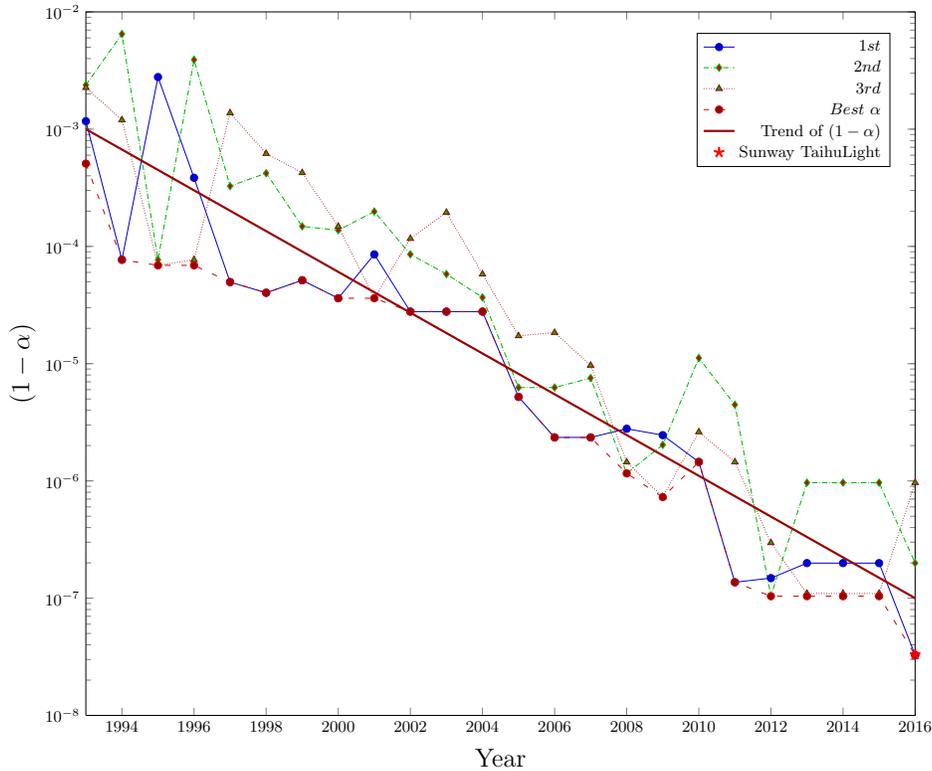
\begin{figure*}
\maxsizebox{1.55\columnwidth}{\textheight}
{
\begin{tabular}{cc}
\begin{tikzpicture}
\begin{axis}
[
	title={\Large Supercomputers, Top 500 1st-3rd},
	width=\textwidth,
	cycle list name={my color list},
		legend style={
			cells={anchor=east},
			legend pos={north east},
		},
		xmin=1993, xmax=2016,
		ymin=1e-8, ymax=1e-2, 
		xlabel=\Large Year,
		/pgf/number format/1000 sep={},
		ylabel=\Large $(1-\alpha)$,
		ymode=log,
		log basis x=2,
		]
\addplot table [x=a, y=b, col sep=comma] {Top500-0.csv};
		\addlegendentry{$1st $}
\addplot table [x=a, y=c, col sep=comma] {Top500-0.csv};
		\addlegendentry{$2nd $}
\addplot table [x=a, y=d, col sep=comma] {Top500-0.csv};
		\addlegendentry{$3rd $}
\addplot table [x=a, y=e, col sep=comma] {Top500-0.csv};
		\addlegendentry{$Best\ \alpha$}
		\addplot[ very thick, color=webbrown] plot coordinates {
			(1993, 1e-3)  
			(2016,1e-7) 
		};
		\addlegendentry{Trend of $(1-\alpha)$}
		\addplot[only marks, color=red, mark=star,  mark size=3, very thick] plot coordinates {
			(2016,33e-9) 
		};
		\addlegendentry{Sunway TaihuLight}
\end{axis}
\end{tikzpicture}
&
\\
\end{tabular}
} 
\caption{Timeline of supercomputer parallelism. The diagrams show ($1-\alpha$) values for the actual first three out of the Top500 
supercomputers over the past 24 years, and to guide the eye, their tendency}
\label{SupercomputerTimeline}
\end{figure*}

\begin{figure*}
\maxsizebox{\textwidth}{\textheight}
{
\begin{tabular}{cc}
\includegraphics{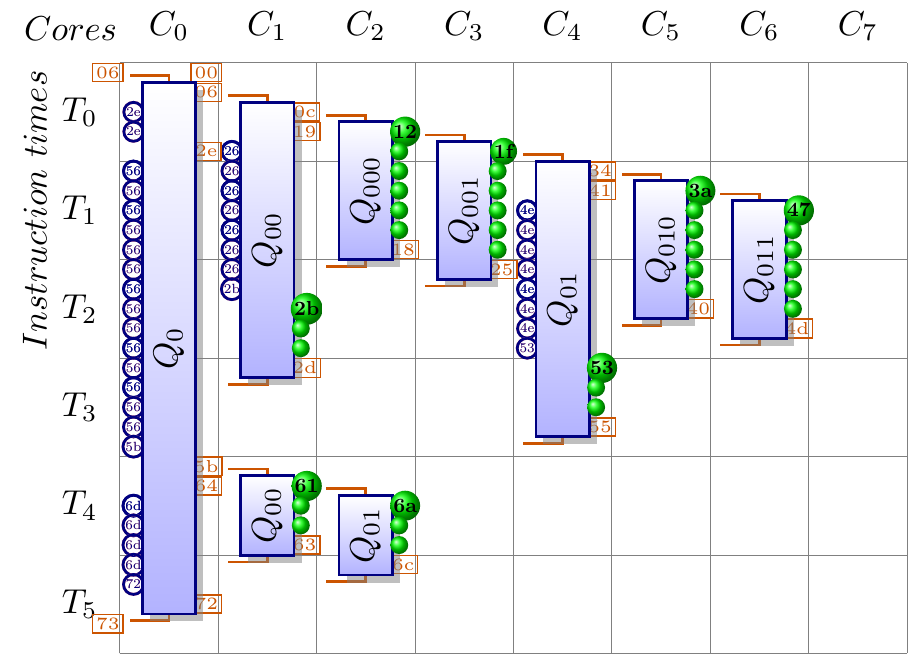}
&
\includegraphics{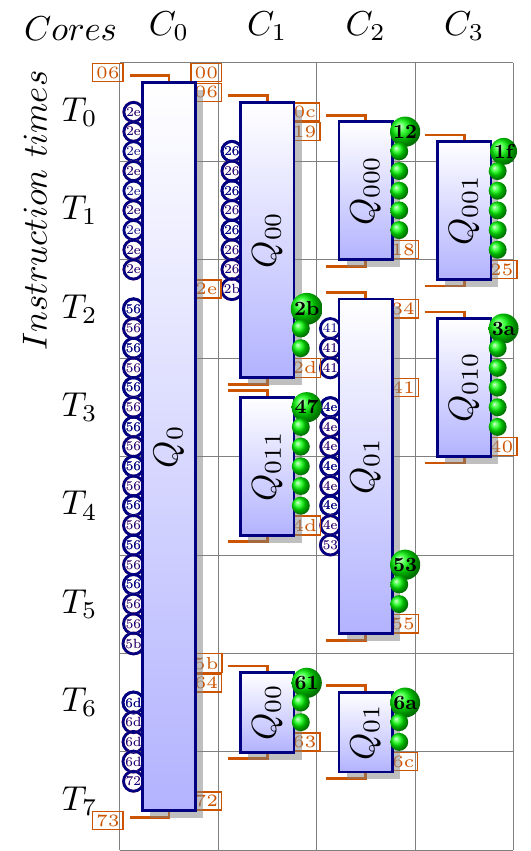}
\\
\end{tabular}
} 
\caption{
The processing graphs corresponding to Figure 1, running on an 8-core (left) and 4-core(right) EMPA processor.}
\label{fig:DynPar}
\end{figure*}

This unusual kind of architecture must respond to
some unusual requirements.
First of all, the architecture must be able to organize itself as the received task requires it,
 and build the corresponding "processing graph", see Fig.~\ref{fig:DynPar}, for legend see \cite{VeghDynamicParallelism:2016}.
Furthermore, it must provide a mechanism for mapping the virtually infinite number of processing nodes to the finite number of cores.
Cores 
\tikz \node[state,scale=\corescale] {$L_{xy}$}; 
must receive the address of the operand, i.e. at least some information must be passed to the rented core.
Similarly, the loaded operand must be returned to the renting core in a synchronized way.
In the first case synchronization is not a problem: the rented core begins its independent life when it receives its operands.
In the second case the rented core finishes its assigned operation and sends the result asyncronously,
independently of the needs of the renting core.
This means that the architecture must provide a mechanism for transferring some (limited amount of) data between cores,
a signalization mechanism for renting and returning cores,
as well as a latched intermediate data storage for passing data
in a synchronized way.

The empty circles are the theoretically needed operations,
and the shaded ones are additional operations of the "smart" cores.
The number of the cores being used changes continuously as they are rented and returned. 
Although \textit{physically} they may be the same core, \textit{logically} they are brand new. Note that the "smart" operations are much shorter -- they comprise simple bit manipulations and multiplexing --,
than the conventional ones that comprise complex machine instructions, and since the rented cores work in parallel (or at least mostly overlap),
the calculation is carried out in 3 clock periods.
The cycle period is somewhat longer, but the attainable parallelism 
approaches the theoretically possible one, and is more than twice as high as
the one attainable  using either two-issue or dual-core processors.

Although the average need of cores is about 3, these cores can be
the simplest processors, i.e. the decreasing complexity of the cores
(over)compensates for the increasing complexity of the processor.
In addition, as the control part of the processors increases,
the need for the hidden parallelization (like out-of-order and speculation) can be replaced by the functionality of the 
flexible architecture, the calculational complexity can be decreased,
and as a result, the clock speed can be increased.
A processor with such an internal architecture appears to the
external world as a "superprocessor", having several times greater
performance than could be extracted from a single-threaded processor.
That processor can adapt itself to the task: unlike in the two issue processor,
all (rented) units are permanently used.
\textit{The many-core systems with flexible architecture comprising cooperating cores can approach
the theoretically possible maximum parallelism.} In addition, the number of the cores can be kept at a strict minimum, allowing reduction of the power consumption.

\section{How long can the parallelism of the many-many processor 
supercomputers still be enhanced, at a reasonable cost?}\label{sec:manymany}

In the  many-many processor (supercomputer) systems the processing units
are assembled using the SPA~\cite{AmdahlSingleProcessor67}, and so their 
maximum performance is bounded by Amdahl's law.
Although Amdahl's original model~\cite{AmdahlSingleProcessor67} is pretty outdated, its simple and clean
interpretation allows us to derive meaningful results even for today's
computing systems. Amdahl assumed that in some $\alpha$ part of the total time
the computing system engages in parallelized activity, in the remaining ($1-\alpha$) part it performs some (from the point of view of parallelization) non-payload activity,
like sequential processing, networking delay, control or organizational operation, etc.
The essential point here is that  all these latter activities behave \textit{as if they were sequential processing}.
Under such conditions, the efficiency $E$ is calculated as the ratio of the
total speedup $S$ and the number of processors $k$:
\begin{equation}
 E = \frac{S}{k}=\frac{1}{k(1-\alpha)+\alpha}\label{eq:soverk}
 \end{equation}

Although in the case of supercomputers ($1-\alpha$) comprises contributions
of a technically different nature (it can be considered as the "imperfectness" of implementation of the supercomputer),
it also behaves as if it were a sequentially processed code.

Fig.~\ref{SupercomputerTimeline} shows how this "imperfectness" was
decreased during the development of supercomputers,
calculated from the actual data of the first three supercomputers in the year in question over a quarter of a century.
As the figure shows, this parameter  behaves similarly to the Moore-observation,
but it is independent of that one
(because the parameter is calculated from $\frac{R_{peak}}{R_{max}}$,
any technology dependence is removed). 

At first glance, it seems to be at least surprising to look for any
dependence in function of "imperfectness". The key is Equ.~(\ref{eq:soverk}).
Since the $\alpha$ approaches unity, the term $k(1-\alpha)$ determines 
the overall efficiency of the computing system. To \textit{increase} $k$ by an order 
or magnitude alone is useless if not accompanied by an order of magnitude \textit{decrease} in the value of ($1-\alpha$). However, while increasing $k$
is simply a linear function, decreasing ($1-\alpha$) as any kind of increasing perfectness, is exponentially more difficult.

Fig.~\ref{SupercomputerTimeline}  proves that today's supercomputers
are built in SPA, and makes it questionable
whether further significant decrease of value ($1-\alpha$) could be reached at reasonable cost.
This means that it is hopeless \textit{to build exa-scale computers, using the principles
drawn from the SPA}.

Looking carefully at $k(1-\alpha)$, one can notice that the two terms
describe two important behavioral features of the computing system.
As already discussed, $(1-\alpha)$ decribes, how much the work
of the many-processor system is \textit{coordinated}.
The factor $k$, on the other hand, describes, how much the processing 
units \textit{cooperate}. In the case of using the SPA, 
the processing units are segregated entities, i.e. they do not cooperate at all.

If we could make a system where the processing units behave differently
in the presence of another processors, we could write $f(k)$ in Equ.~(\ref{eq:soverk}).
Depending on how cores behave together
in the presence of another cores when solving a computing task, the $f(k)$,
the cooperation of the processing units can drastically increase
the efficiency of the many-processor systems.
In other words, to increase the performance of many-many-processor computers,
\textit{the cores must cooperate} (at least with some) other cores. \textit{Using  cooperating cores is inevitable for building supercomputers at a reasonable cost.}

\section{Can we eliminate non-payload calculations 
by replacing them with architectural changes?
}\label{sec:manymany}

A computer computes everything, because it cannot do any 
other type of operations. Computational density has reached its
upper bound, so no further performance increase in that direction is possible. In addition to introducing different forms of HW and SW
parallelism, it is possible to omit some non-payload, do-not-care calculations, through providing and utilizing special HW signals instead.
The  signals can be provided for the participating cores, and can be used
to replace typical calculational instruction sequences by using special hardware signals.
The compilation is simple: where the compiler should generate 
non-payload loop organization commands, it should give a hint about renting a core
for executing non-payload instructions and providing external 
synchronization signals.

A simple example: when summing up elements of a vector, the only 
payload instruction is the respective \lstinline|add|. 
One has, however, to address the operand (which includes handling the index,
calculating the offset and adding it to the base address), to advance the 
loop counter, to compare it to the loop bound, and to jump back conditionally. All those non-payload operations can be replaced 
by handling HW signals, if the cores can cooperate, resulting
in a speed gain of about 3, using an extra core only.
Even, since the intermediate sum is also a do-not-care value
until the summing is finished, a different sumup method can be used,
which may utilize dozens of cores and result in a speed gain
of dozens.  When organizing a loop,
the partial sum is one of the operands, so it must be read before
adding a new summand, and must be written back to its temporary storage,
wasting instructions and memory cycles; in addition it excludes 
the possibility of parallelizing the sumup operation.  
For details and examples see~\cite{VeghDynamicParallelism:2016}.
 
This latter example also demonstrates that \textit{the machine instruction 
is a too rigid atomic unit of processing}.
\textit{Utilizing HW signals from cooperating cores rather than providing some conditions 
through (otherwise don-not-care) calculations, allows us to eliminate 
obsolete calculational instructions, and thus apparently
accelerate the computation by a factor of about ten.}

\section{Do we really need to pay with an indeterministic operation
for multiprocessing?}\label{sec:multiproc}

The need for multi-processing (among others) forced to use exceptional instruction
execution. I.e., a running process is \textit{interrupt}ed, its HW and SW state
is saved and restored, because the hard and soft parts of the \textit{only} processor must be lent to another process.
The code of the interrupting process is effectively inserted in the flow of executing the interrupted code.
This maneuver causes an indeterministic behavior of the processor:
the time when two consecutive machine instructions in a code flow are executed,
becoming indeterminate.  

The above is due to the fact that during development, some of the really successful
accelerators, like the internal registers and the highest level cache,
became part of the architecture: the soft part of the processor.
In order to change to a new thread, the current soft part must be saved in
(and later restored from) the memory. Utilizing asynchronous interrupts as well as operating system services,
 implies a transition to new operating mode, which is a complex and very time-consuming process.

All these extensions were first developed when the computer systems had 
only one processor, and the only way to provide the illusion of running several
processes, each having its own processor, was to detach the soft part 
from the hard one. Because of the lack of proper hardware support, this 
illusion depended on using SW services and on the architectures being constructed 
with a SPA in mind, conditions that require rather expensive 
execution time: in modern systems a context change may require
several thousands of clock cycles. As the hyper-threading proved,
detaching soft and hard part of the processors results in considerable performance enhancement.

By having more than one processor and the Explicitly Many-Processor Approach~\cite{VeghDynamicParallelism:2016},
the context change can be greatly symplified.
For the new task, such as providing operating system services and servicing
external interrupts a dedicated core can be reserved. The dedicated core  can be prepared and held in supervisor mode. When the execution
of the instruction flow follows, it is enough to clone the relevant part
of the soft part: for interrupt servicing nothing is needed, for using OS services only
the relevant registers and maybe cache.
(The idea is somewhat similar to utilizing shadow registers for servicing an
asynchronous interrupt.)

If the processors can communicate among each other using HW signals
rather than OS actions, and some communication mechanism, different from using
(shared) memory is employed, the apparent performance of the computing systems
becomes much faster. \textit{For cooperating cores no machine instructions (that waste real time, machine and
memory cycles) are needed for a context change, allowing for a several hundredfold more rapid
execution in these spots.}
The application can even run parallel with the system code, allowing further
(apparent) speedup.

Using the many-processor approach creates many advantageous changes in the
real-time behavior of the computing systems. Since the processing units
do not need to save or restore anything, the servicing can start immediately
and is restricted to the actual payload instructions.
The dedicated processing units cannot be addressed by non-legal
processing units, so issues like exluding priority inversion are handled
at HW level. And so on.

\section{The common part: implement supervised cooperating cores, handling extra signals 
and storages}\label{sec:manymany}

From all points of view (the just-a-few and many-many processors, as well as utilizing kernel-mode or real-time services) we arrive at the 
same conclusion: segregated processors in the many-processor systems
do not allow a greater increase in the performance of our computing systems,
while cooperating processors can increase the attainable single-threaded performance. Amdahl contented this by  a half century ago: 
"\emph{
	the organization of a single computer has reached its limits and that truly significant advances can be made only by interconnection of a multiplicity of computers in such a manner as to permit cooperative solution.}" \cite{AmdahlSingleProcessor67}

At this point the many-core architectures have the advantage that they are
in the close proximity to one another: there is no essential difference
between that a core needing to reach its own register (or signal) or  that of another core.
The obstacle is actually the SPA: for a core
and a process, there exists no other core.

In the suggested new approach, which can be called  \textit{Explicitly
Many-Processor Approach} (EMPA), the 
cores (through their supervisor) can know 
about their neighbours. 
Today, radical departures from conventional approaches 
(including rethinking the complete computing stack) are advanced  \cite{Esmaeilzadeh:2015:AAP:2830689.2830693},
but at the same time a smooth transition must be provided to that
radically new technology.
\textit{To preserve compatibility with conventional computing, 
the EMPA approach~\cite{VeghDynamicParallelism:2016} is phrased using the terms of conventional 
computing} (i.e. it contains SPA as a subset).

\section{How do algorithms benefit from the EMPA architecture?}

Some of the above-mentioned boosting principles are already implemented
in the system. From the statistics one can see that in some spots,
performance gain in the range 3-30 can be reached. 
The different algorithms need different new accelerator 
building stone solutions in frame of EMPA.

For example,
the gain 3 in an executing loop, when used in an image processing task
where for edge detection a 2-dimensional matrix is utilized, means nearly an order of magnitude performance gain, using the same calculational  architecture in calculating a new point. And, to consider all
points of the picture another double loop is used. This means, that 
a 4-core EMPA processor can produce nearly 100 times more rapid processing (not considering that several points can be processed in parallel on processors with more cores).
This is achieved not by increasing computing density,
but by replacing certain non-payload calculations with HW signals,
and so executing 100 times less machine instructions.

\section{How Amdahl's dream can be implemented?}\label{sec:dream}

The MC architecture comprising segregated cores is indeed broken. It can, however, be mended, if the manycore chips are manufactured in the form using cooperating cores.

As the first step toward implementing such a system, for simulating its sophisticated 
internal operation and providing tools for understanding and validating it,
an EMPA development system~\cite{VeghDynamicParallelism:2016} has been prepared.
An extended assembler prepares EMPA-aware object code, while  the simulator
allows us to watch the internal operation of the EMPA processor.

To illustrate the execution of programs using the EMPA method, 
a processing diagram is automatically prepared by the system,
and different statistics are assembled.
Fig.~\ref{fig:DynPar} shows the equivalent of Fig.~\ref{fig:flexibleproc},
running on an 8-core and a 4-core processor,
respectively (for legend see ~\cite{VeghDynamicParallelism:2016}).
The left hand figure depicts the case when "unlimited" number of processing units are available,
the right hand one shows the case when the processor has a limited number of computing resources to implement
the maximum possible parallelism.

The code assembled by the compiler is the same in both cases. The supervisor logic
detects if not enough cores are available (see right side), and 
delays the execution (outsourcing more code) of the program fragments until some cores are free again.
The execution time gets longer if the processor cannot rent enough cores
for the processing, but the same code will run in both cases,
without deadlock and violating dependencies.

 For electronic implementation, some ideas may be borrowed
from the technology of reconfigurable systems. There, in order to minimize
the need for transferring data, some local storage (block-RAM) is located
between the logical blocks, and a LOT of wires is available for connecting them.

In analogy also with FPGAs, the cores can be implemented as mostly fixed
functionality processing units, having multiplexed connecting wires to
their supervisor with fixed routing.
Some latch registers and non-stored program functionality gates 
can be placed near those blocks, which can be accessed by both cores and supervisor. 
The inter-core latch data can be reached from the cores using
pseudo-registers (i.e. they have a register address, but are not part of the register file)
and the functionality of the cores also depends on the inter-core signals.
In the prefetch stage the cores can inform the supervisor about the presence
of metainstruction in their object code, and in this way the mixed code instructions can be directed to the right destination.
In order to be able to organize execution graphs, the cores
(after renting) are in parent-child relation to unlimited depth. 

As was very correctly stated~\cite{Vishkin:BrokenManycoreCACM}, "due to its high level
of risk, prototype development fits best
within the research community."
The principles and practice of EMPA differ radically from those of SPA. To compare the performance of both, EMPA needs a range of development. Many of the present components, accelerators, compilers, etc., 
with SPA in mind, do not fit EMPA.
The research community can accept (or reject) the idea,
but it definitely warrants some cooperative work.

\bibliographystyle{abbrv} 

\end{document}